\documentstyle[prl,aps]{revtex}

\begin{document}
\draft
\title{Skyrmions in disordered heterostructures}
\author{A.J. Nederveen and Yu. V. Nazarov}
\address{Faculty of Applied Physics and DIMES, Delft University of Technology,
Lorentzweg 1, 2628 CJ Delft,\\
The Netherlands}
\date{\today}
\maketitle

\begin{abstract}
We have investigated the effect of weak disorder on the ground state of a
two dimensional electron gas in the quantum Hall regime at filling factors
slightly deviating from unity. The skyrmions and antiskyrmions are 
found to be present even at filling factor $\nu=1$. They may be strongly
squeezed by the disorder. We have studied temperature effect on
skyrmion density and size.
\end{abstract}


Recently it has been proved that the lowest energy charged excitations of a
two-dimensional electron gas (2DEG) in quantum Hall regime at filling factor 
$\nu =1$ are skyrmions\cite{sondhi}. These excitations involve
many electrons and can be
adequately described in the framework of a semiclassical $\sigma $-model
as the
topologically non-trivial  spin textures of an isotropic ferromagnet
\cite{belavin}. Owing to the incompressibility of the
Quantum Hall ground state the topological charge of a skyrmion is equal
to its
electric charge\cite{macdonald}.

There is a convincing experimental evidence for these collective
excitations.\cite{exp}. However, the experiments show only
 qualitative agreement with the theory\cite{schmeller,maude}. 
It has been suggested \cite{schmeller,maude} 
that the agreement will be better if the effects of disorder are
taken into account. Despite the importance of the topic, very few
theoretical studies \cite{Green,Lilly} deal with skyrmions in disordered heterostuctures.
The weak disorder limit has received no attention.

In this letter we show that even a weak disorder gives rise to 
(anti)skyrmions even at filling factor $\nu =1$. 
In the limit of a small Zeeman energy the Skyrmion radius is decreased 
by the disorder potential. We present the results for
the (anti)skyrmion density at filling factors deviating from unity.
We have found an interesting  temperature effect
on skyrmion
density and radius. We attempted to compare our results with  
experimental data of \cite{maude}.

In the limit of small $\tilde{g}$, which stands for the ratio between the
Zeeman and exchange energy, skyrmions can be described with an effective
energy functional derived in\cite{sondhi}. The energy of a single skyrmion
as a function of its radius $a$ reads\cite{sondhi,yuli} 
\begin{equation}
E_{\text{sk}}=E_{0}+E_{\text{Zeeman}}+E_{\text{Coulomb}}=\frac{1}{4}\sqrt{%
\frac{\pi }{2}}\frac{e^{2}}{\varepsilon l_{B}}+\frac{g\mu _{B}Ba^{2}}{%
l_{B}^{2}}\ln \left( \frac{r_{s}}{a}\right) +\frac{3\pi ^{2}}{64}\frac{e^{2}%
}{\varepsilon a}  \label{skyrmionenergy}
\end{equation}
where $E_{0}$ is the skyrmion gap energy, $l_{B}$ is the magnetic length, $%
g $ is the electron $g$ factor,  $r_{s}\simeq
l_{B}\tilde{g}^{1/2}$. Although the gap energy dominates the total
energy, the radius is set by the two smaller terms depending  on $a$. The
equilibrium Skyrmion radius $a_{0}$ can be found by minimizing the total
energy with respect to $a$. In this article we  neglect the slowly changing
logarithmic factor  in the Zeeman energy and incorporate it to $g$.

The disorder is brought into a 2DEG  mainly by the donors situated in a layer
several hundreds of angstroms away from the electron layer. We incorporate
the effect of disorder in a fourth term in the energy functional. For a
Skyrmion at position ${\bf r}_{0}$ in a random donor potential $V({\bf r})$ 
 this energy reads 
\begin{equation}
E_{\text{dis}}({\bf r}_{0})=\int d^{2}r\rho ({\bf r-r}_{0},a)V({\bf r})
\label{impurity}
\end{equation}
where $\rho ({\bf r},a)=(a/r^2+a^2)^2/\pi$ is the charge density of a Skyrmion with 
radius $a$. The potential $V({\bf r})$ is not
screened by the electrons of the 2DEG, since at integer filling
factors the 2DEG is an insulator. 
We assume the Gaussian distribution of $V({\bf r})$ with a correlator
\begin{equation}
\left\langle \left| V_{q}\right| ^{2}\right\rangle =\frac{U_{0}^{2}}{%
q^{2}}  \label{correlator}
\end{equation}
which is calculated by averaging over donor
positions.   
Weak disorder requires $U_{0} \ll E_{0}$. 
If we assume that donor positions do not correlate, we come up
with an unrealistically big estimate of $U_{0}\simeq e^{2}/\varepsilon d$,  $d$
being the distance between the 2DEG and the donor plane. In realistic
heterostructures the observed disorder is known to be much weaker, possibly due to
correlations between donor positions. Therefore, we assume that $U_{0}\ll
e^{2}/d,$ $E_{0}$. $U_{0}$ can be extracted from the experimentally measured
mobility at zero magnetic field.

At a very qualitative level, our results are straightforward.
At zero temperature it is energetically favorable for a Skyrmion to exist 
in the potential wells which are sufficiently deep (Fig. 1), 
so that $E_{\text{dis}}<E_{\text{0}}$. 
Since the  disorder is weak, such wells are infrequent, so
that the Skyrmion density will be much smaller than the electron density.
These deep wells are to be expected very steep, that compresses the
skyrmion to make it fit into the well. (see Fig. 1) This agrees with
the preliminary results of computer simulations reported in
\cite{Lilly}.

At temperatures of the order of $U_0$ the skyrmions may appear 
in the wells which are less deep (left side of Fig.1), that increases their density. 
Since the potential
minima become less steep, the skyrmion radius  becomes bigger. 
The same consideration is valid also
antiskyrmions which are situated
in the maxima of the random potential.

Let us make these simple ideas quantitative.
To calculate the skyrmion density at filling factor $\nu =1$ and zero
temperature, we consider the following integral 
\begin{equation}
n_{\text{sk}}(E,a)={\int d^{2}r\sum\limits_{m}\delta ({\bf r}-{\bf r}%
_{m})\delta (a-a_{m})\delta (E_{\text{dis}}({\bf r},a)-E)}/{\int d^{2}r}
\label{dichtheidintegraal}
\end{equation}
where $n_{\text{sk}}(E,a)$ is the density of skyrmions having  potential $E_{%
\text{dis}}=E$ with radius $a$ per interval of energy and radius (that we can call 
skyrmion {\it density of states}) and ${\bf \{r}_{m},a_{m}\}$ is a set of Skyrmion positions and radii for a certain 
$V({\bf r})$-configuration. These ${\bf \{r}_{m},a_{m}\}$ are not arbitrary, but
determined from the fact that the skyrmions are in equilibrium. Since we assume
that Skyrmion density is small, we can consider isolated Skyrmions. It means
that we disregard interactions between Skyrmions when they are in different
wells, but due to strong repulsive interaction within the well there is only
{\it one Skyrmion} in each well. In this case a Skyrmion only interacts with the
disorder potential. This results in three conditions, two for Skyrmion
position: $\partial _{x}E_{\text{dis}}=\partial _{y}E_{\text{dis}}=0$ and
one for Skyrmion radius: $\partial _{a}E_{\text{dis}}=-\partial _{a}E_{\text{%
sk}}$. The actual density of skyrmions is to be found by integrating
$n_{\text{sk}}(E,a)$ over all radii and $E<E_{\text{sk}}\approx E_0$.

To proceed we rewrite this integral as follows. To incorporate the equilibrium
conditions we mentioned above, we transform $\delta $-functions 
in terms of the three equilibrium conditions.
%follows
%\begin{equation}
%\sum\limits_{m}\delta ({\bf r}-{\bf r}_{m})\delta (a-a_{m})\rightarrow
%\delta (\partial _{x}E_{\text{dis}})\delta (\partial _{y}E_{\text{dis}%
%})\delta (\partial _{a}E_{\text{sk}}+\partial _{a}E_{\text{dis}})
%\label{transform}
%\end{equation}
The Jacobian of this transformation enters the integral: $\left| \partial
_{\alpha \beta }(E_{\text{sk}}+E_{\text{dis}})\right| $, with $\alpha ,\beta
=\left\{ x,y,a\right\} $. 
Them we average Eq. (\ref{dichtheidintegraal}) at ${\bf r}=0$ over all possible $V_{q}$%
-configurations. We have to take into account that skyrmions are situated in
minima, so that that all eigenvalues of the Jacobian are positive. 
%This yields 
%\begin{equation}
%\begin{array}{ll}
%n_{\text{sk}}(E,a)= & \left\langle \delta (\partial _{x}E_{\text{dis}%
%})\delta (\partial _{y}E_{\text{dis}})\delta (\partial _{a}E_{\text{sk}%
%}+\partial _{a}E_{\text{dis}})\times \left| \partial _{\alpha \beta }(E_{%
%\text{sk}}+E_{\text{dis}})\right| \right. \\ 
%& \left. \times \theta (\partial _{xx}E_{\text{dis}})\theta (\partial
%_{xx}E_{\text{dis}}\partial _{yy}E_{\text{dis}}-\left( \partial _{xy}E_{%
%\text{dis}}\right) ^{2})\theta (\left| \partial _{\alpha \beta }(E_{\text{sk}%
%}+E_{\text{dis}})\right| )\right\rangle _{V_{q}}
%\end{array}
%\label{dichtheid2}
%\end{equation}
 We express $E_{\text{dis}%
}$ in terms of Fourier components: $E_{\text{dis}}(0)=\sum\limits_{q}E_{\text{dis}%
}(q) $, with $E_{\text{dis}}(q)=V_{q}\rho _{q}$ and $\rho _{q}=aqK_{1}\left(
aq\right) $ \cite{sondhi,belavin}. By adding extra variables we can rewrite
all $\delta $- and $\theta $-functions in the form of exponents. Then
integration over the random potential appears to be Gaussian and can be
performed. In the course of  averaging the terms
like $\sum\limits_{q}\rho _{q}^{2} <V_q^2>$ will appear. Formally they diverge
at $q\rightarrow 0$. To deal with this divergence we have to recall that
the interaction between Skyrmions is a long range one. So that it becomes
effectively important at large distances and Skyrmions screen out components
of disorder potential having $q<\sqrt{n_{\text{sk}}}$, inverse distance
between Skyrmions. So we cut off integration at $q\simeq \sqrt{n_{\text{sk}}}$. 

 After integration over extra variables  the
expression for density of states reads 
\begin{equation}
\begin{array}{ll}
n_{\text{sk}}(E,a)= & 
%TCIMACRO{
%\dfrac{2^{15}\pi ^{10}a^{3}}{\sqrt{3}(4\alpha -3)^{\frac{7}{2}}U_{0}^{4}} }
%BeginExpansion
{\displaystyle {2^{15}\pi ^{10}a^{3} \over \sqrt{3}(4\alpha -3)^{\frac{7}{2}}U_{0}^{4}}}%
%EndExpansion
\left( -%
%TCIMACRO{\dfrac{4}{3} }
%BeginExpansion
{\displaystyle {4 \over 3}}%
%EndExpansion
%TCIMACRO{\dfrac{E}{a^{2}} }
%BeginExpansion
{\displaystyle {E \over a^{2}}}%
%EndExpansion
+%
%TCIMACRO{\dfrac{\partial _{a}E_{\text{sk}}}{a} }
%BeginExpansion
{\displaystyle {\partial _{a}E_{\text{sk}} \over a}}%
%EndExpansion
\left( 4\alpha -1\right) +\partial _{a}^{2}E_{\text{sk}}\left( 4\alpha
-3\right) \right) \\ 
& \left( 
%TCIMACRO{\dfrac{\partial _{a}E_{\text{sk}}}{a} }
%BeginExpansion
{\displaystyle {\partial _{a}E_{\text{sk}} \over a}}%
%EndExpansion
\left( 3-6\alpha \right) +%
%TCIMACRO{\dfrac{E}{a^{2}} }
%BeginExpansion
{\displaystyle {E \over a^{2}}}%
%EndExpansion
\right) ^{2}\exp \left( -24\pi 
%TCIMACRO{
%\dfrac{\frac{1}{3}E^{2}-E\partial _{a}E_{\text{sk}}a+\left( \partial _{a}E_{\text{sk}}\right) ^{2}a^{2}\alpha }{U_{0}^{2}\left( 4\alpha -3\right) } }
%BeginExpansion
{\displaystyle {\frac{1}{3}E^{2}-E\partial _{a}E_{\text{sk}}a+\left( \partial _{a}E_{\text{sk}}\right) ^{2}a^{2}\alpha  \over U_{0}^{2}\left( 4\alpha -3\right) }}%
%EndExpansion
\right)
\end{array}
\label{dichtheid3}
\end{equation}
where 
\begin{equation}
\alpha =2\pi \sum_{q}%
%TCIMACRO{\dfrac{\rho _{q}^{2}}{q^{2}} }
%BeginExpansion
{\displaystyle {\rho _{q}^{2} \over q^{2}}}%
%EndExpansion
\simeq -\ln (a\sqrt{n_{\text{sk}}})  \label{alpha}
\end{equation}
Here we take into account  that the sum in Eq. \ref{alpha} 
diverges and must be cut off at $q\simeq \sqrt{n_{%
\text{sk}}}$. We assume that $a\sqrt{n_{\text{sk}}}\ll 1$. So that $\alpha$
depends on Skyrmion concentration and eqs.(\ref{dichtheid3}),(\ref{alpha}) form together a self consistency problem.

Now we evaluate the radius $a$ which minimizes the exponent of eq. (\ref
{dichtheid3}). We substitute the Skyrmion energy of eq. (\ref{skyrmionenergy}%
) in the exponent of eq. (\ref{dichtheid3}). It appears that there are two
limits corresponding to the strength of the Zeeman energy

\begin{equation}
a_{\text{opt}}=\left\{ 
\begin{array}{lllll}
%TCIMACRO{\dfrac{3\pi ^{2}}{2^{7}} }
%BeginExpansion
({\displaystyle {3\pi ^{2} \over 2^{7} \tilde g}})^{-\frac{1}{3}}%
%EndExpansion
l_{B} & = & a_{0}, &  & \tilde{g}^{\frac{1}{3}}%
%TCIMACRO{\dfrac{E_{0}}{U_{0}} }
%BeginExpansion
{\displaystyle {E_{0} \over U_{0}}}%
%EndExpansion
\gg 1 \\ 
%TCIMACRO{\dfrac{3\pi ^{\frac{3}{2}}}{2^{\frac{5}{2}}} }
%BeginExpansion
{\displaystyle {3\pi ^{\frac{3}{2}} \over 2^{\frac{5}{2}}}}%
%EndExpansion
\alpha  l_{B} & \ll & a_{0}, &  & \tilde{g}^{\frac{1}{3}}%
%TCIMACRO{\dfrac{E_{0}}{U_{0}} }
%BeginExpansion
{\displaystyle {E_{0} \over U_{0}}}%
%EndExpansion
\ll 1
\end{array}
\right.  \label{straal}
\end{equation}
where $a_{0}$ is the skyrmion radius in the absence of a random potential
field.

Now we can proceed with calculation of the total density.
We expand the exponent around $a=a_{\text{max}}$ and $E=-E_{\text{sk}}$.
We restrict ourselves to the more interesting limit of small
Zeeman energy: $\tilde{g}^{\frac{1}{3}}\frac{E_{0}}{U_{0}}\ll 1$. In this
limit we encounter again two different limits depending on the strength of
the Zeeman energy. The point is that the prefactor, which is in fact the determinant $%
\left| \partial _{\alpha \beta }(E_{\text{sk}}+E_{\text{dis}})\right| $,
vanishes at $a=a_{opt}$ in the first order approximation. In
both cases the exponent is the same, 
$n_{\text{sk}} \propto \exp ( 2\pi E^2_0/\alpha)$.  
This is sufficient to solve the self consistency problem.  
The leading approximation to $\alpha$ is
$\alpha _{0}=\sqrt{\pi }\frac{E_{0}}{U_{0}}$, 
the next order approximation is necessary to evaluate the
prefactor. Finally we obtain for
the density 
\begin{equation}
n_{0}=\left\{ 
\begin{array}{lll}
\gamma _{1}\left( 
%TCIMACRO{\dfrac{U_{0}}{E_{0}} }
%BeginExpansion
{\displaystyle {U_{0} \over E_{0}}}%
%EndExpansion
\right) ^{\frac{9}{4}}n_{e}\exp \left( -2\sqrt{\pi }%
%TCIMACRO{\dfrac{E_{0}}{U_{0}} }
%BeginExpansion
{\displaystyle {E_{0} \over U_{0}}}%
%EndExpansion
\right) , &  & \tilde{g}^{\frac{1}{3}}%
%TCIMACRO{\dfrac{E_{0}}{U_{0}} }
%BeginExpansion
{\displaystyle {E_{0} \over U_{0}}}%
%EndExpansion
\ll \left( 
%TCIMACRO{\dfrac{U_{0}}{E_{0}} }
%BeginExpansion
{\displaystyle {U_{0} \over E_{0}}}%
%EndExpansion
\right) ^{\frac{1}{3}} \\ 
\gamma _{2}\tilde{g}\left( 
%TCIMACRO{\dfrac{E_{0}}{U_{0}} }
%BeginExpansion
{\displaystyle {E_{0} \over U_{0}}}%
%EndExpansion
\right) ^{4}\left( 
%TCIMACRO{\dfrac{U_{0}}{E_{0}} }
%BeginExpansion
{\displaystyle {U_{0} \over E_{0}}}%
%EndExpansion
\right) ^{\frac{9}{4}}n_{e}\exp \left( -2\sqrt{\pi }%
%TCIMACRO{\dfrac{E_{0}}{U_{0}} }
%BeginExpansion
{\displaystyle {E_{0} \over U_{0}}}%
%EndExpansion
\right) , &  & \tilde{g}^{\frac{1}{3}}%
%TCIMACRO{\dfrac{E_{0}}{U_{0}} }
%BeginExpansion
{\displaystyle {E_{0} \over U_{0}}}%
%EndExpansion
\gg \left( 
%TCIMACRO{\dfrac{U_{0}}{E_{0}} }
%BeginExpansion
{\displaystyle {U_{0} \over E_{0}}}%
%EndExpansion
\right) ^{\frac{1}{3}}
\end{array}
\right.  \label{dichtheid4}
\end{equation}
where $n_{e}$ is the {\it electron} density in the 2DEG, $\gamma_1
\simeq 9 \times 10^{-4}$, $\gamma_2 \simeq 6 \times 10^{3}$ are numerical factors.
Here we have taken into account that the number of skyrmions at
$\nu=1$ is equal to the number of antiskyrmions, and those two equally
contribute to $n_{sk}$.

Therefore we find three regimes in dependence of the relative strength of
the Zeeman energy with respect to disorder. The most probable Skyrmion radius is equal to
its disorderless value at $\tilde{g}^{\frac{1}{3}}\frac{E_{0}}{U_{0}}\gg 1$
and to $a_{opt} \simeq \frac{9 \pi^2}{4} \sqrt{\frac{E_0}{ 2 U_0}} l_B \ll a_0$
in the opposite limit. 
At $\tilde{g}^{\frac{1}{3}}\frac{E_{0}}{U_{0}}\simeq \left( \frac{U_{0}}{E_{0}}\right) ^{\frac{1}{3}}$
the prefactor changes.

Let us evaluate  skyrmion densities
at filling factors slightly deviating from unity. In the absence of
disorder, skyrmion and antiskyrmion densities are
proportional to the filling factor deviation (Fig. 2),
$n_{\text{sk,ask}}= \pm \delta f \theta(\pm \delta f)$.
where $\delta f=n_{e}(\nu-1)$. This is no longer correct in the presence of disorder potential.
We note that at unity filling factor the
chemical potential lies precisely in the middle of the gap between skyrmion and antiskyrmion states. 
If the filling factor deviates from unity, we have to
recalculate the densities for a shifted chemical potential $\mu$. 
Fortunately, this is simple. The density of states exhibits exponential
dependence on energy, $n_{sk}(E) \propto \exp(-E/\Omega)$, $\Omega \equiv 
2 \alpha U_0^2/E_0$. 
This is why the skyrmion density is changed by a factor $\exp (-\mu
/\Omega )$ whereas the 
antiskyrmion one by $\exp (\mu /\Omega )$. 
An attention should be paid to the fact that the change of the chemical potential
also affects $\alpha$. We note that the total charge per unit area must correspond to the filling
factor, that is $n_{\text{sk}}-n_{\text{ask}}= \delta f$. This allows as to express $\mu$
 in terms of $\delta f$ and obtain for the densities
\begin{equation}
\left\{ 
\begin{array}{l}
n_{\text{sk}} \\ 
n_{\text{ask}}
\end{array}
\right\} =\sqrt{%
%TCIMACRO{\dfrac{\delta f^{2}}{4} }
%BeginExpansion
{\displaystyle {\delta f^{2} \over 4}}%
%EndExpansion
+ 
%TCIMACRO{
%\dfrac{n_{0}^{4}}{\delta f^{2} +\sqrt{\delta f^{4} +16 {n_{0}^{4}} }
%BeginExpansion
{\displaystyle {n_{0}^{4} \over {\displaystyle {\delta f^{2}}}+\sqrt{{\displaystyle {\delta f^{4}}}+16 n_0^4}}}%
%EndExpansion
}\pm 
%TCIMACRO{\dfrac{\delta f}{2} }
%BeginExpansion
{\displaystyle {\delta f \over 2}}%
%EndExpansion
\label{dichtheid6}
\end{equation}

These densities are plotted in Fig. 2, together with the ratio of the total
number of spin flips $n_{sk}+n_{ask}$ and the number of spin flips at 
$\delta f =0$. The latter quantity can be observed experimentally
since it determines the rounding of the spin polarization
peak.\cite{exp}

Let us consider the effect of a finite
temperature. Still we have to assume 
that the Skyrmion density is much smaller than the electron
density, that is, $T \ll E_0$. Also the temperature should be low enough
for Coulomb interaction to make it
impossible for two Skyrmions to occupy  the same potential well. 
Under these circumstance the skyrmions
effectively behave as fermions, so we can make use of Fermi
statistics.\cite{statistics}
To calculate the total density at $\mu =0$ we integrate 
 the density of states  of Eq. (\ref{dichtheid3}) 
multiplied by the Fermi distribution function, $n_{\text{sk}}(E,a)\times
f(E-E_{0})$(see Fig. 3 a,b).
If we expand the exponent in $n_{\text{sk}}(E)$ near $E=E_{\text{sk}}$ 
yielding a term linear in $E-E_{\text{sk}}$ we see that the integral
converges only at $T<T_c$,$T_{c}={U_{0}}/{4\sqrt{\pi }}$. 

Of course, it does not mean that the density actually diverges
at $T=T_c$. Rather, it indicates an exponential temperature
dependence of the density at $T>T_c$. To reveal this one,
 we go back
to the expression (\ref{dichtheid3}) and consider the most important
term in the exponent which is quadratic in energy. This term competes 
with the exponent of  the Fermi
distrubution tail. 
%We shall minimize  the exponent 
%\begin{equation}
%-8\pi 
%%TCIMACRO{\dfrac{E^{2}}{U_{0}^{2}(4\alpha -3)} }
%%BeginExpansion
%{\displaystyle {E^{2} \over U_{0}^{2}(4\alpha -3)}}%
%%EndExpansion
%+%
%%TCIMACRO{\dfrac{-E-E_{0}}{k_{B}T} }
%%BeginExpansion
%{\displaystyle {-E-E_{0} \over k_{B}T}}%
%%EndExpansion
%\label{exponent2}
%\end{equation}
The minimum is achieved at $E=E_{min}= \alpha T_c^2 T$.
The solving  the self consistency problem gives  
$\alpha(T)= E_0/(2T+2T_c^2/T)$. 
%\begin{equation}
%\alpha(T) = 
%%TCIMACRO{
%%\dfrac{E_{0}}{2\T \left( 1+\dfrac{T_c^2}{T^{2}}\right) } }
%%BeginExpansion
%{\displaystyle {E_{0} \over 2T \left( 1+{\displaystyle {T_c^{2} \over T^{2}}}\right) }}%
%EndExpansion
%\label{alpha2}
%\end{equation}
 At the critical temperature  $E_{\text{min}}=E_{0}$, 
as it should be. 
The density at $T>T_c$ thus reads
\begin{equation}
n(T) \propto \exp(-E_{0}%
%TCIMACRO{\dfrac{T }{T ^{2}+T_{c}^{2}} }
%BeginExpansion
{\displaystyle {T  \over T^{2}+T_{c}^{2}}})%
%EndExpansion
\label{exponent3}
\end{equation}

and exhibits a non-Arrenius behavior. The logarithm of
the density is plotted in Fig.3 versus inverse temperature.
The optimal size of a (anti)skyrmion can be extracted from the exponent
(\ref{dichtheid3}). At $T>T_c$ and it grows with temperature: $a(T)
=a(0) T/T_c$, the latter is valid till $a(T) \ll a_0$. This is because
the potential minima become less steep at higher energies.

With our method, we can also obtain the finite temperature results for $\nu \ne 1$.

Our results can be checked with spin polarization measurements.
Indeed, the finite density of (anti)skyrmions at $\nu=1$ would manifest
itself as a rounding of the spin polarization peak.
The reduction of their size $a$ due to disorder and its restoration at $T>T_c$
can be detected as a change of skyrmion spin $\propto a^2$.
However, this check requires a very accurate measurement of the spin polarization peak
in the close vicinity of $\nu=1$ at different temperatures, that has not
been performed yet.
 
As to transport measurements, we attempted to compare our results with experimental data of 
the Ref. \cite{maude}. We assume the longitudinal
resistivity to be proportional to the skyrmion density: $\rho _{xx}\sim n(T)$
and the temperature dependence of $n(T)$ to be dominated by the exponent (\ref{exponent3}). 
In Fig. 3 the experimental data are fit to the theoretical curve, 
with fit parameters $U_{0}\simeq 15$ K and $%
E_{0}\simeq 12$ K. Although $U_0 \simeq E_0$, we expect our theory
to be qualitatively true since the numerical factors provide big
exponents. Indeed, in this case $T_{c}\simeq 2 \text{K} \ll E_0$.
From the given mobility we extract $U_0 \simeq 10 \text{K}$.

However, most of the transport measurements \cite{schmeller} exhibit no saturation of
$\rho_{xx}$ down to 1 K. Although this can be explained by better
quality of the heterostructures ($U_0 < 3 \text{K}$), we hesitate to
make a point out of our fit. 
The point is that all transport measurements, whatever interpreted, give an estimation
of $E_0$ which is an order of magnitude smaller than the theoretical value. 
Possibly this indicates that the assumption $\rho _{xx}\sim n(T)$ is no
good. Our results explain how this may happen. Indeed, even at zero temperature
and $\nu=1$ we have a finite concentration of (anti)skyrmions. 
What may be thermally activated is their {\it mobility} rather than
concentration. It is no surprise since they have to overcome high
potential barriers when moving between potential minima.

In conclusion, we have developed the theory of skyrmions in weakly
disordered heterostructures. The skyrmions appeared to be present
at filling factor $\nu =1$. The disorder and temperature strongly affect
their density and size.

The authors are indebted to G. E. W. Bauer, H. T. C. Stoof, and S. E.
Korshunov for interesting discussions.
This work is supported bye the "Stichting voor
Fundamenteel Onderzoek der Materie"~(FOM), and 
 the "Nederlandse Organisatie voor Wetenschappelijk Onderzoek"
~(NWO).

\begin{figure}
\caption
{Skyrmions in disorder potential. Their size must be small to fit into a
deep well.}
\label{fig1}
\end{figure}
\begin{figure}
\caption
{Skyrmion densities versus filling factor.
}
\label{fig2}
\end{figure}
\begin{figure}
\caption
{
Temperature dependence of the skyrmion density.
Solid line presents our theoretical results, the 
squares correspond to the fitted data of Ref. 6.
}
\label{fig3}
\end{figure}

\begin{references}
\bibitem{statistics} We stress that this is effective statistics that
has nothing in common with the actual quantum mechanical statistics
of free skyrmions. The latter is not relevant here.
\end{references}
\end{document}